\documentclass[a4paper,11pt,reqno,superscriptaddress]{revtex4}

\usepackage[centertags]{amsmath}
\usepackage{amsfonts}
\usepackage{amssymb}
\usepackage{amsthm}
\usepackage{mathrsfs}
\usepackage{newlfont}
\usepackage{stmaryrd}
\usepackage{mathrsfs}
\usepackage{euscript}
\usepackage{graphicx}
\usepackage{enumerate}
\usepackage{tikz}
\usepackage{pgf}
\usetikzlibrary{positioning,fit,calc}
\usetikzlibrary{positioning}
\usetikzlibrary{arrows,automata}
\usepackage{wrapfig}
\usepackage{subfigure}
\usepackage{amscd}
\usepackage{hyperref}

\theoremstyle{plain}

\theoremstyle{definition}

\theoremstyle{remark}

 \let\be=\beta  
\let\ve=\varepsilon  \let\ga=\gamma 
\let\ka=\kappa \let\la=\lambda  
\let\si=\sigma

 \let\Ga=\Gamma  \let\Om=\Omega

\DeclareMathAlphabet{\mathpzc}{OT1}{pzc}{m}{it}

 \let\be=\beta  
\let\ve=\varepsilon  \let\ga=\gamma 
\let\ka=\kappa \let\la=\lambda  
\let\si=\sigma

 \let\Ga=\Gamma  \let\Om=\Omega

\newcommand{\caF}{{\mathcal F}}

\newcommand{\bbI}{{\mathbb I}}

\newcommand{\opunit}{\text{1}\kern-0.22em\text{l}}

\newcommand{\rel}{\,|\,}

\newcommand{\id}{\textrm{d}}

\def\dbar{\,{\mathchar'26\mkern-12mu \text{d}}}

\def\bea{\begin{eqnarray}}
\def\eea{\end{eqnarray}}
\def\ba{\begin{array}}
\def\ea{\end{array}}
\def\n{\nonumber}

\begin{document}

\title{Statistical forces from close-to-equilibrium media}

\author{Urna Basu}
%\email{christian.maes@fys.kuleuven.be}
%\affiliation{SISSA, Trieste, Italy and Instituut voor Theoretische Fysica, KU Leuven, Belgium}
\affiliation{SISSA, Trieste, Italy} 
\affiliation{Instituut voor Theoretische Fysica, KU Leuven, Belgium}
\author{Christian Maes}
\affiliation{Instituut voor Theoretische Fysica, KU Leuven, Belgium}
\author{Karel Neto\v{c}n\'{y}}
\email{netocny@fzu.cz}
\affiliation{Institute of Physics, Academy of Sciences of the Czech Republic, Prague, Czech Republic}

\begin{abstract}
  We discuss the physical meaning and significance of statistical forces on quasi-static probes in first order around detailed balance for driven media. Exploiting the quasi-static energetics and the structure of (McLennan) steady nonequilibrium ensembles, we find that
  the statistical force obtains a nonequilibrium correction deriving from  the excess work of driving forces on the medium
  in its relaxation after probe displacement. 
  This reformulates, within   a more general context, the recent result by N. Nakagawa (Phys.~Rev.~E \textbf{90}, 022108 (2014)) on thermodynamic aspects of weakly nonequilibrium adiabatic pumping. It also proposes a possible operational tool for accessing some excess quantities in steady state thermodynamics.  Furthermore, we show that the point attractors of a (macroscopic) probe coupled to a weakly driven medium realize the predictions of the minimum entropy production principle. Finally, we suggest a method to measure the relative dynamical activity through different transition channels, via the measurement of the statistical force induced by a suitable driving. 
\end{abstract}
\maketitle

Statistical forces are responsible in thermodynamics for generating transport of energy, momentum or matter as a result of the irreversible tendency to approach equilibrium \cite{Groot}. They can be realized as true mechanical forces by coupling a probe to the macroscopic medium. The probe can itself be a macroscopic device like a wall or a piston with pressure as the statistical force. Another example are elastic forces which can be thought of as entropic forces when all interactions are ignored, working simply by the power of large numbers \cite{entrspring}.  For our set-up (Fig.\ref{fig:setup}(a)) we have in mind a dilute suspension of colloids (= probe particles) in a fluid (= medium) with mutual coupling, i.e., both colloid and fluid react to each other as dictated from  an interaction potential. We assume however that the colloid is quasi-static, meaning that its characteristic time is much longer than that of the fluid. 
The resulting effective dynamics of the colloid picks up various aspects of the fluid; there are the friction and the noise as usual for motion in a thermal bath, but because of our assumption of infinite time-scale separation we concentrate here exclusively on the systematic force which is the statistical average over the fluid degrees of freedom of the mechanical force on the colloid; see \cite{2nd,stef} for further discussion on friction and noise in nonequilibrium media. The general question concerning thermodynamics of active or driven media is of much current interest, e.g. for exploring the validity of equations of state in nonequilibrium \cite{kafri, Solon, Matthias, kardar}.

For a probe in contact with an equilibrium reservoir the free energy is a potential for statistical forces.  The present paper studies these forces for reservoirs that are subject to weak driving.  By the latter we mean that the fluid particles are undergoing rotational (nonconservative) forces with dissipation in yet another background environment that will just be represented by its temperature; see~Fig.\ref{fig:setup}(a).  The main question is to see how that nonequilibrium feature corrects the gradient
statistical force derived from the (equilibrium) free energy.  Or, {\it vice versa}, how the force on the colloid teaches us about irreversible thermodynamic features of the fluid.  The result is that to linear order in the amplitude of the rotational forces the work on the probe equals the excess work done on the fluid by the rotational forces  in its relaxation to the new stationary condition corresponding to the slightly displaced probe.  A similar result was already obtained in~\cite{naoko14}
in the context of cyclic adiabatic pumping.

Excess quantities are omnipresent in discussions on steady state thermodynamics.  Their origin is theoretical, trying to distinguish steady state effects from transient effects also for nonequilibrium fluids.  Indeed when driven, the fluid obtains a stationary dissipation with some mean entropy production rate in the environment. % (or possibly internal degrees of freedom).
However, as some external parameters change in time, relaxational processes of the nonequilibrium fluid will also contribute to (excess) dissipation.  The origin of such decomposition, housekeeping {\it versus} excess dissipation, is probably found in the work of Glansdorff and Prigogine \cite{gp,jsp}, but it has since been repeatedly stressed also in more recent studies of steady state thermodynamics \cite{oono,HatanoSasa,sasatasaki,clausius,cla}.  For example, for thermal properties of nonequilibrium systems one introduces the excess heat which defines nonequilibrium heat capacities \cite{epl,cejp}.  One recurrent difficulty however is to find a good operational meaning of these excess quantities.  Nature does not dissipate the steady heat and the excess heat separately; similar for the notion of excess work.  That is why it can be useful to find that  the statistical force on a probe is \emph{directly} related to excess work, at least close to equilibrium and for thermodynamic transformations controlled by mechanical motion of a probe.

A further motivation of the present work is to complete the close-to-equilibrium theory of steady state thermodynamics with the nature of statistical forces.  Clearly and as we will see in Section \ref{linr} statistical forces enter in the First Law for the energy balance.  They are therefore very much part of the theory of irreversible thermodynamics for composed systems (here, probe plus fluid).  Moreover, as is the content of Section \ref{min}, the question appears in what sense these statistical forces realize the minimum entropy production principle; see \cite{scholar}.  In other words, whether we can understand statistical forces as the way in which systems achieve minimum entropy production rate.  The answer is positive in the sense that indeed the very requirement of minimal entropy production rate for the composed system again and also determines the statistical force in terms of the excess work.

A third direction in which statistical forces are interesting, is that they are able to make visible aspects of
(time-symmetric) dynamical activity.  That is not surprising because excess work involves the dynamics and hence, in contrast with the free energy which is static, kinetic factors will be present in the statistical force.  We build that into a ``frenometer'' to get explicit information about the relative dynamical activity through reactivity channels; see Section \ref{2ch}.

We begin the paper with a thermodynamic approach based on specifying the energy balance close-to-equilibrium.  We find the relation between excess work of the medium, the force done on the probe and the nonequilibrium heat capacity.  Section~\ref{sm} gives the corresponding statistical mechanical basis.   We need the McLennan ensemble theory to determine the correction to the equilibrium statistical forces.  It gives a second derivation of the result that relates excess work of the medium with the work to displace the probe.
We end Section~\ref{sm} with a discussion about the validity of
our result  when kinematical time reversal is included (like for underdamped diffusions). Section~\ref{line} is devoted to a detailed illustration of the framework in context of a linear system.
The relation between excess work and statistical forces is rederived using the minimum entropy production principle in Section~\ref{min}. The relation between statistical work and relative dynamical activity is contained in Section~\ref{2ch}, suggesting as we already mentioned, a simple
``frenometer''.

The present work follows and substantially extends~\cite{PRL-attempt} where the main idea has been reported.

% --------------------------------------------------------------
\section{Energetics of irreversible thermodynamics}\label{linr}
We refer to Fig.~\ref{fig:setup}(a) for a cartoon of three classes of particles.  There is the probe on which a force is induced by its contact with a medium and a heat bath.  The medium is subject to nonequilibrium conditions and dissipates into the (equilibrium) heat bath at temperature T.
In general $x$ denotes the ``position'' (possibly multi-dimensional) of the probe.  With $f$ the statistical force, the corresponding work performed by moving the probe is $f\cdot \id x$.  The stationary energy of the medium when the probe is at $x$ is denoted by $E(x)$.  Then, the quasi-static energetics (or ``First law'') of the nonequilibrium medium is generally given by the balance equation for the energy as
\begin{equation}
\label{bal}
\id E(x) = -f\cdot \id x + \dbar W^\text{ex}(x) +
\dbar Q^\text{ex}(x)
\end{equation}
where
$\dbar W^\text{ex}$ denotes the excess thermodynamic work of the driving forces in the medium along the relaxation process that corresponds to the thermodynamic transformation
$x\rightarrow x+\id x$, $T\rightarrow T + \id T$, and similarly
$\dbar Q^\text{ex}$ is the (incoming) excess heat.
Note that we speak about excesses because the stationary medium constantly dissipates work into heat; excess is the extra corresponding to the transient process of reaching a new stationary condition.
We assume that the excess heat satisfies a Clausius relation  $\dbar Q^\text{ex} = T \id S(x)$ with $T$ the temperature and
$S(x)$ can then be called the calorimetric entropy.  We do not need its detailed expression here.
The assumption can be checked (as we do in the next Section~\ref{sm}) in the linear regime around thermodynamic equilibrium;
the original proof is found in the work of
Komatsu \emph{et ~al.}~\cite{cla,prl08}.
We define the free energy
${\cal F}(x) = E(x) - T S(x)$ for which then, {\it cf.}~also~\cite{cla},
\begin{equation}
\id {\cal F} = -S\,\id T - f \cdot \id x + \dbar W^\text{ex}
\end{equation}
By the (equilibrium) minimum free energy principle we know  that there is no linear order correction in
$\caF$ or $\id \caF$, meaning that in the considered linear regime
${\cal F}(x)$ coincides with the equilibrium free energy
${\cal F}_\text{eq}(x) = E_\text{eq}(x) - TS_\text{eq}$ where the First Law for equilibrium combined with the Clausius equality is $\id E_\text{eq}(x) = -f_\text{eq}\cdot \id x + T\id S_\text{eq}$.
Expanding around equilibrium,
$f = f_\text{eq} + g$, $S = S_\text{eq} + \tilde s$, the first-order contributions yield zero free energy change and hence,
within the first-order approximation,
\begin{equation}\label{lin-irr}
\dbar W^\text{ex} \simeq \tilde s\,\id T + g \cdot \id x
\end{equation}
In particular for isothermal processes
($\id T = 0$), we find
\begin{equation}
\label{ge}
g\cdot \id x \simeq \dbar W^\text{ex}  %,\qquad \id T=0
\end{equation}  for the nonequilibrium (to first order around equilibrium) component of statistical force in terms of the excess work, whereas for
$\id x = 0$ the excess work is related to the nonequilibrium entropy correction $\tilde s$, which is itself related
to the nonequilibrium heat capacity \cite{epl,cejp}.

Our observation on the absence of the first-order correction in the free energy provides a simple variation of
formula~(13) in~\cite{naoko14} by Nakagawa for the work transfer during cyclic adiabatic pumping in terms of nonequilibrium (excess) heat into the driven system.  However, we do not restrict ourselves
to any specific protocol of operation.
Formula \eqref{ge} gives a direct relation between the
mechanical force on the probe on the slow time scale and the
steady-state thermodynamic
process in the medium on the fast time scale. Remark that the excess
quantities, though omnipresent
in steady state thermodynamics, \emph{cf.}~the balance equation \eqref{bal}, are known to be not easily accessible
\emph{directly}. Hence, formula (4) could be used to access some of the
excess quantities in a mechanical way.

In the next section we give the statistical mechanical basis for the above general thermodynamic arguments.

\section{Statistical mechanical approach}\label{sm}

We closely follow the approach of Komatsu, Nakagawa, Sasa and Tasaki~\cite{clausiusjap}.
Yet we start from a general set-up which formalizes the idea of statistical force on quasi-static probes. There will be no need to introduce or indeed to specify the time-evolution except that we assume in general that the medium to which the probe is coupled passes through stationary states of some generic (McLennan) form.

We think of $\eta$ as the collection of degrees of freedom of a driven medium. For the rest of the paper we assume these variables are even under kinematic time-reversal, so not containing velocity degrees of freedom as for example with underdamped diffusions; the results do not change however in the more general case --- see Section \ref{addr}. For simple convenience we take them discrete so that we use sums when computing averages etc.  The medium particles undergo rotational forces of order $\varepsilon$ and they obtain a stationary regime by dissipating heat into a thermal bath at temperature $T$; we also write
$\be = T^{-1}$ setting Boltzmann's constant $k_B = 1$.   Each stationary regime of the medium depends on the position $x$ of a slow probe. The probe is immersed in the medium and the contact is modeled via a joint interaction potential $U(x,\eta)$
which by assumption also includes the interaction among the medium particles as well as the self-interaction of the probe if present.
As the medium is supposed to be macroscopic it is relevant to define the statistical force on the probe as the average mechanical force,
\begin{equation}\label{def}
f(x) = - \sum_\eta \rho_x(\eta) \, \nabla_x U(x,\eta) = -\langle \nabla_x U(x,\eta) \rangle^x
\end{equation}
where the average is over the steady nonequilibrium stationary distribution $\rho_x$ of the $\eta-$medium at fixed $x$.  Note that the total system is composed (medium particles plus probe) but we work under the hypothesis that the $\eta-$variables are relaxing much faster.  When we apply that to the case of an equilibrium medium we find the standard result that the statistical force is given as the gradient of the free energy.  In statistical mechanical writing that free energy is
${\cal F}_\text{eq}(x)=
-T\log Z_x$ with $Z_x$ the equilibrium partition function corresponding to the $\eta-$medium when in equilibrium with fixed probe position $x$: the distribution is then  given by the Boltzmann-Gibbs factor $\rho_x^{\text{eq}}(\eta) = \exp \{-\beta U(x,\eta)\}/Z_x$.

To go beyond equilibrium, we need information about $\rho_x$ for determining \eqref{def}.  We work under the condition of local detailed balance for the nonequilibrium medium~\cite{tas,mn,KLS,leb} which relates the probe--medium dynamics with the entropy fluxes into the  environment (= a heat bath at temperature $T$). By our assumption the driving forces breaking the global detailed balance provide a contribution to the entropy fluxes proportional to some small parameter $\ve$.

\begin{figure}[ht]
 \centering
 \includegraphics[width=6 cm]{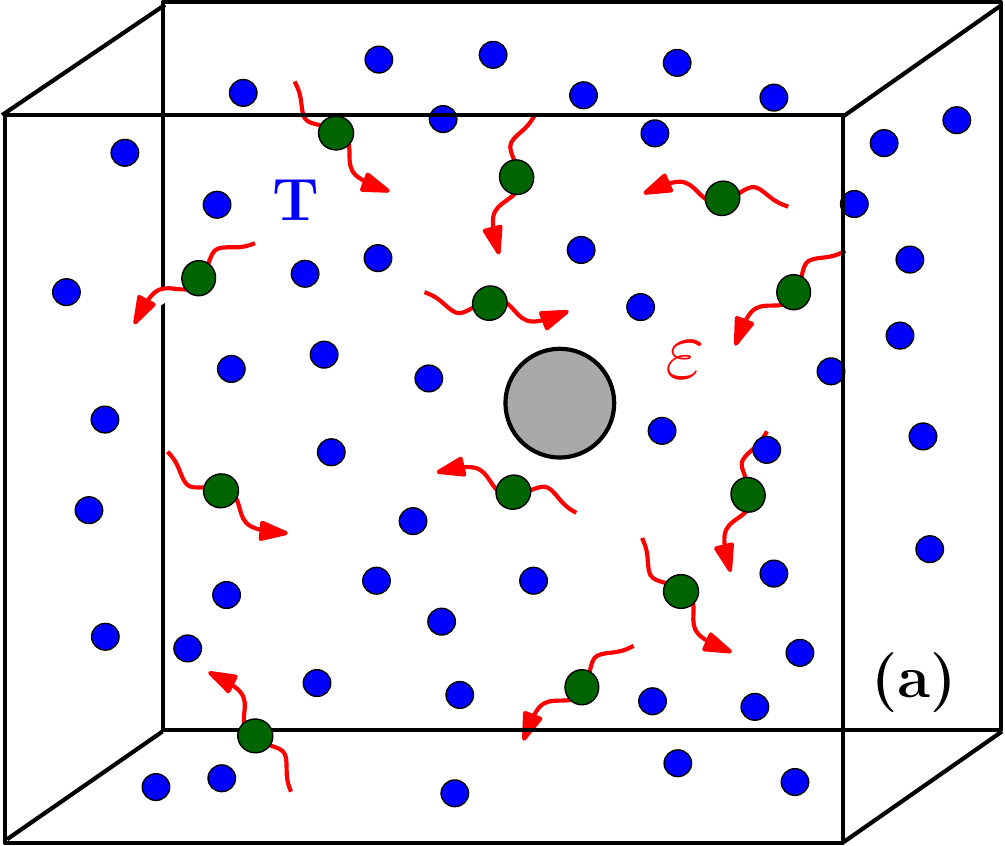}\hspace*{1.2 cm} \includegraphics[width=6 cm]{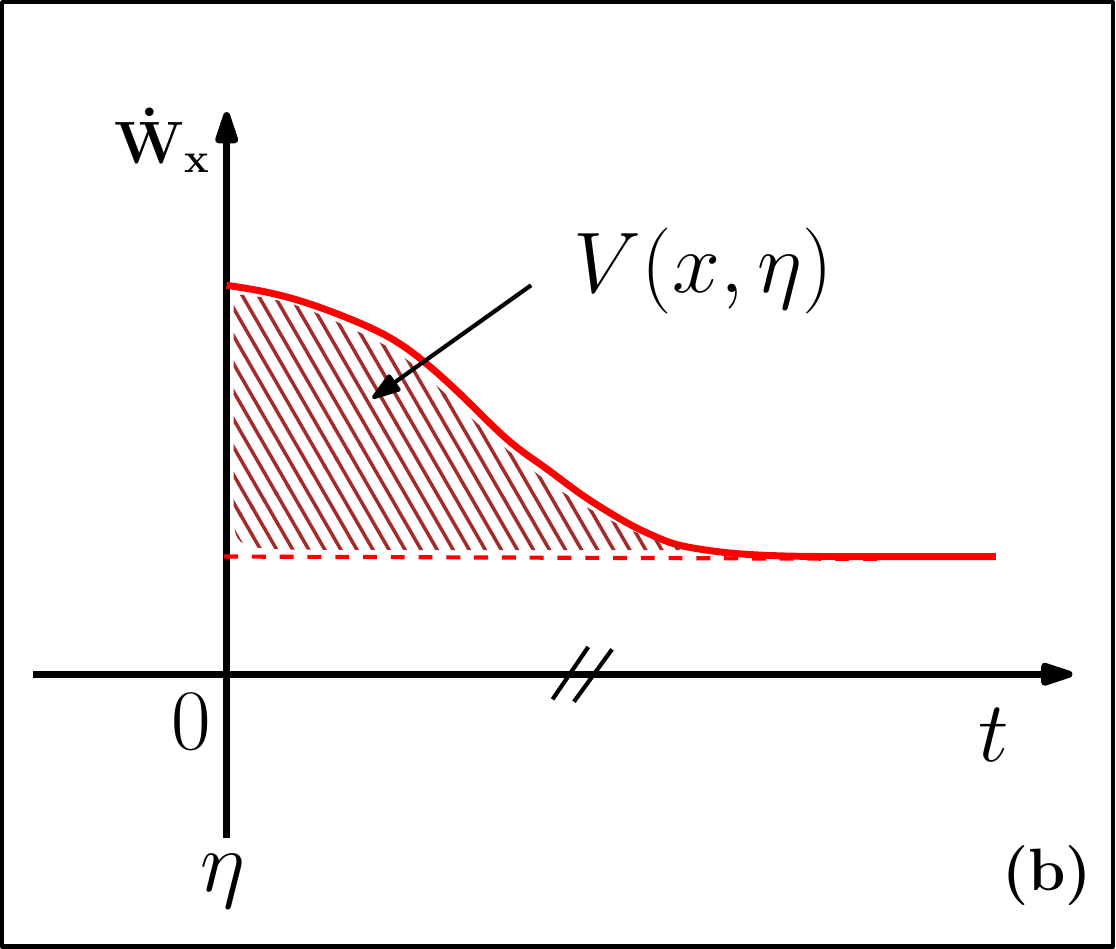}
 % probe_sea.pdf: 595x842 pixel, 72dpi, 20.99x29.70 cm, bb=0 0 595 842
 \caption{(a) A slow probe (light grey disc) is immersed in a nonequilibrium medium (green arrowed circles), in contact with an equilibrium reservoir (small blue circles). (b) Excess work $V(x,\eta)$ done by the driving forces  in relaxing to the stationary condition for a fixed probe postion $x$ starting from medium configuration $\eta.$
 Here $\dot W_x$ denotes the mean instantaneous power of the driving forces. }
 \label{fig:setup}
\end{figure}

Close to equilibrium the medium is well described by  the McLennan stationary ensemble \cite{mcl,mac}
\begin{equation}\label{ml}
\rho_x^{\text{ML}}(\eta) = \frac{1}{{\cal Z}_x}\,e^{-\be(U + V)(x,\eta)}
\end{equation}
Here $V(x,\eta)$ is the excess work of driving forces along the relaxation process started from $\eta$ with $x$ fixed having zero expectation $\langle V\rangle^x =0$ under the stationary distribution $\rho_x$; see Fig.~\ref{fig:setup}(b) and \eqref{defV} in Appendix~\ref{amc} for the definition.  Note that $V$ is itself of order $\varepsilon$.  It turns out that
\begin{equation}\label{corr}
\rho_x = \rho_x^{\text{ML}} + O(\varepsilon^2),\quad {\cal Z}_x = Z_x + O(\varepsilon^2)
\end{equation}
with $Z_x$ the %unmodified
equilibrium partition function (at $\varepsilon=0$).  Formula~\eqref{ml} describes the steady  linear regime around equilibrium. For example, linear response formul{\ae} can be derived from it; see \cite{mcl}.

Specific examples follow below in Sections \ref{line} and \ref{2ch}.

% ---------------------------------------------------------
\subsection{Deriving the energy balance}\label{sec:energy}

The stationary energy $E(x) = \langle U\rangle^x$ changes as
\[
\id \langle U\rangle^x = \langle \id U\rangle^x + \sum_\eta \id\rho_x(\eta)\,U(x,\eta)
\]
The medium is doing work $f\cdot \id x$ on the probe, hence
\begin{equation}
\langle \id U \rangle^x = -f \cdot \id x
\end{equation}
mimicking \eqref{def} as the mechanical energy $U$ does not depend on temperature.
The excess work when the medium relaxes from the stationarity under $x$ to the new stationarity under $x+\id x$ reads
\begin{equation}
\dbar W^\text{ex}(x) = \sum_\eta \rho_x(\eta)\,V(x + \id x,\eta) =
\langle \id V \rangle^x
\end{equation}
where we have used $\langle V\rangle^x = 0$. Using that condition again and writing
$\langle \id V \rangle^x = -\sum_\eta \id\rho_x(\eta)\,V(x,\eta)$,
the (renormalized) First law~\eqref{bal} is verified by defining the excess heat as
\begin{equation}\label{wex}
\dbar Q^\text{ex}(x) =
\sum_\eta \id \rho_x(\eta)\,(U + V)(x,\eta)
\end{equation}

Let us now use the statistical mechanical formul{\ae} \eqref{ml}--\eqref{corr} to work in the linear regime, and use $\rho_x\rightarrow \rho_x^{\text{ML}}$ to replace the stationary distribution in leading order around equilibrium.
The Clausius equality
$\dbar Q^\text{ex} = T\,\id S(x) + O(\varepsilon^2)$ with entropy
$S(x) = -\sum_\eta \rho_x(\eta)\,\log\rho_x(\eta)$, can be obtained directly from the definitions when using the McLennan distribution
\eqref{ml}.  The free energy equals
\begin{equation}
\label{fre}
{\cal F}(x) = \langle U \rangle^x- T S(x) = -T\log Z_x + O(\varepsilon^2)
\end{equation}
and indeed has no linear order correction.  That verifies the hypotheses involved in the thermodynamic derivation of \eqref{lin-irr}.  We can however also give a direct derivation inserting \eqref{ml} into \eqref{def}, which
comes next.

\subsection{Excess work equals the nonequilibrium correction to statistical work}
\label{sm-derivation}

When the medium undergoes nonequilibrium driving, there is a new stationary nonequilibrium density,
\begin{equation}\label{hs}
\rho_x(\eta) = \rho_x^{\text{eq}}(\eta)\,[1 + h_x(\eta)]
\end{equation}
in terms of a density $h_x$ (of order $\varepsilon$) with respect to the reference equilibrium distribution.
The equilibrium distribution $\rho_x^{\text{eq}}(\eta)$ satisfies the identity
\bea
\nabla_x \rho_x^{\text{eq}}(\eta) 
&=& - \rho_x^{\text{eq}}(\eta)\,\left[ \frac 1T \nabla_x U(x,\eta)\, + \nabla_x \log Z_x \right] \n 
\eea
We  obtain the statistical force $f(x) = T\nabla_x \log Z_x + g(x) $ by multiplying the above relation with $\rho_x(\eta)/\rho_x^{\text{eq}}(\eta)$ and summing over $\eta.$ The nonequlibrium correction $g(x)$ is then  given in terms of the density $h_x$ defined in \eqref{hs} above,
\begin{equation}\label{11}
g(x) = -T\,\sum_\eta \nabla_x h_x(\eta) \,  \rho_x^{\text{eq}}(\eta) = -T\,\langle \nabla_x h_x\rangle^{x,\text{eq}}
\end{equation}
Since  $\,\sum_\eta h_x(\eta) \,  \rho_x^{\text{eq}}(\eta) = \langle h_x\rangle^{x,\text{eq}} = 0$ (from the normalization applied to \eqref{hs}), for a small displacement $\id x$ of the probe the work done is
\begin{equation}
g(x)\cdot \id x = -T\,\sum_\eta h_{x+\id x}(\eta) \, \rho_x^{\text{eq}}( \eta) = -T \langle h_{x+\id x}\rangle^{x,\text{eq}}
\label{pert}
\end{equation}
When interested in first order around equilibrium we can as well write
\begin{equation} \label{wowa}
g(x)\cdot \id x = -T\,\langle h_{x+\id x}\rangle^x + O(\varepsilon^2)
\end{equation}
with respect to the stationary distribution of the medium in contact with an external thermal bath at temperature $T$.  The McLennan distribution \eqref{ml} gives $h_x(\eta) = -\frac 1T V(x,\eta) + O(\varepsilon^2)$ and combining that with \eqref{wex} we recover \eqref{ge}.

\subsection{Including kinematical time-reversal}\label{addr}

The result that certain excess quantities as encountered in steady state thermodynamics are accessible via mechanical measurements, remains valid in a broader context than considered so far.
We have in mind the case of medium variables $\eta$ containing velocity degrees of freedom or, more generally, dynamical degrees of freedom that are not even under kinematic time-reversal.  We indicate here briefly where some changes in the arguments would occur.\\

First, the purely thermodynamic argument of Section \ref{linr} does not change at all. The entropy $S$ used there will however get a slightly more general statistical mechanical appearance than in Section~\ref{sec:energy}.  We have to use the symmetrized Shannon entropy introduced by Komatsu {\it et al.}; see e.g. \cite{cla} for a recent review.  Calling $\pi$ the kinematic time-reversal (like flipping the sign of all momenta)
and assuming that the equilibrium reference is $\pi-$invariant,
$\rho^\text{eq}_x(\pi\eta) = \rho^\text{eq}_x(\eta)$,
we have the Clausius relation (in first order $\varepsilon$ around equilibrium)
$\dbar Q^\text{ex} = T \id S(x) + O(\varepsilon^2)$ with entropy
$S(x) = -\frac 12\sum_\eta [\rho_x(\eta)+ \rho_x(\pi\eta)]\,\log\rho_x(\eta)$.  That relation again follows by taking for $\rho_x(\eta)$ the McLennan distribution, but the excess work $V(x,\eta)$ does not appear directly in the statistical weight.  Rather, the nonequilibrium correction to the equilibrium distribution has the more general form
$h_x(\eta) = -\frac{1}{T} V(x,\pi\eta) + O(\ve^2)$.
Despite the modification of the entropy function, the formulas~\eqref{pert}--\eqref{wowa} yield
$g(x) \cdot \id x = \langle V(x + \id x,\eta) \rangle^x + O(\ve^2)$ without any change. Since
$\langle V(x + \id x,\eta) \rangle^x = \langle V(x + \id x,\eta) \rangle^{x,\,\text{eq}} + O(\ve^2)$ still equals the excess work
$\dbar W^\text{ex}(x)$ for the transformation $x \mapsto x + \id x$, we have checked that our main relation~\eqref{ge} indeed extends to this more general case.

%--------------------------------------------------------
\section{Linear model}\label{line}

As an illustration we consider as medium a cloud of non-interacting particles driven by linear rotational forces and diffusively moving in a viscous fluid. The linearity is assumed also for the interaction with the probe as well as for a potential force trapping the cloud in a bounded region. It allows exact calculations and is a good approximation for weak nonlinearities.

\begin{figure}[ht]
 \centering
  \includegraphics[width=7.5 cm]{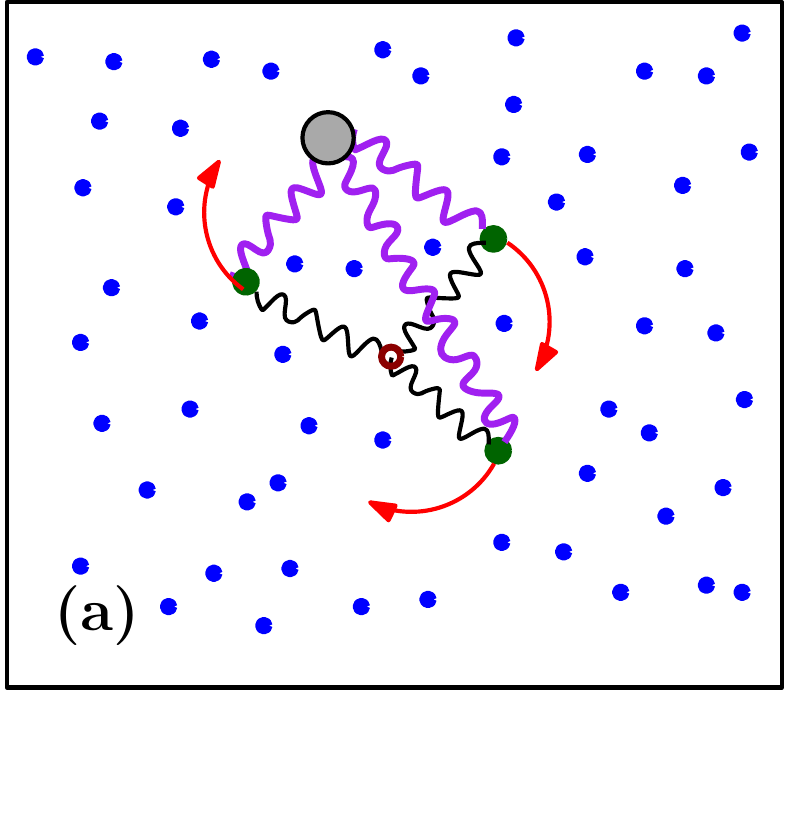} \includegraphics[width=8 cm]{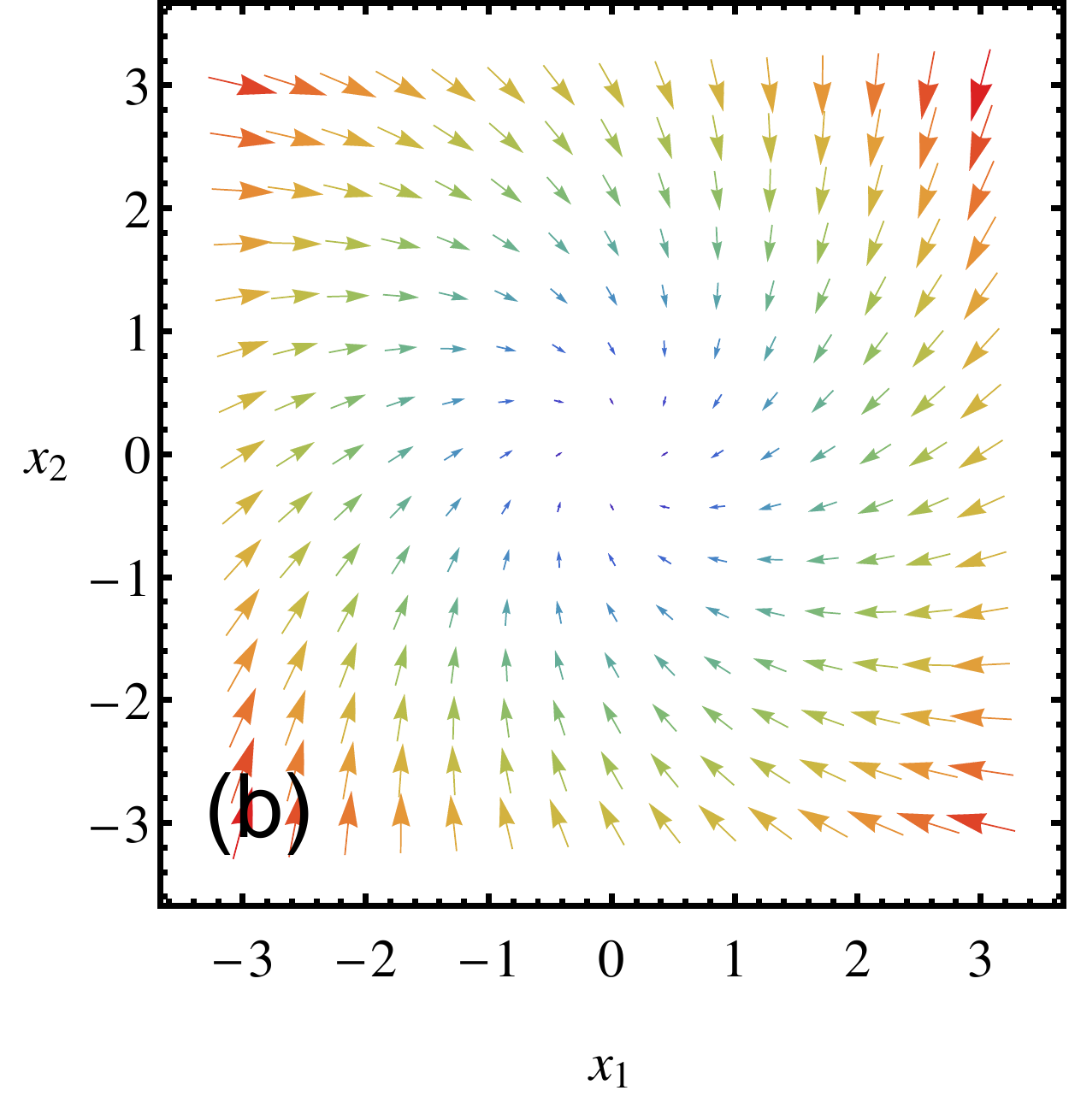}
 % probe_sea.pdf: 595x842 pixel, 72dpi, 20.99x29.70 cm, bb=0 0 595 842
 \caption{The linear model: (a) A slow probe connected with  harmonic springs to the cloud particles which in turn are connected to the origin and driven by a rotational force.
 (b) The statistical force for the 2-dimensional example with nonequilibrium driving $\ve=2.5$ where $\la=5.0$ and $b=0.1.$}\label{fig:linear}
\end{figure}

The cloud consists of many particles from which it will make sense consider statistical average but as we take them independent, it suffices to consider just one of them.  That  generic particle lives in $d$ dimensions with coordinate
$y = (y_1,\ldots,y_d)$; we use here $y$ instead of $\eta$ for better accordance with position degrees of freedom.  See Fig.~\ref{fig:linear}(a) for a $d=2-$representation.

Let $A_s>0$ and $B, K \geq 0$ be symmetric $d\times d-$matrices.  The total potential is
\begin{equation}\label{potential}
U(x,y) = \frac{1}{2}\,y \cdot A_s y + \frac{\lambda}{2}\, (x -  y) \cdot B  (x - y) + \frac{1}{2}\,x \cdot K x
\end{equation}
where the second term is the interaction potential
$U_I(x,y)$ with the probe at position $x$ at coupling strength $\lambda>0;$ the matrix
$K$ stands for the ``bare spring constant'' of the probe.  We also include a rotational force on the cloud described by an arbitrary antisymmetric matrix $A_a$ and with $\ve$ characterizing its magnitude. %See Fig.~\ref{fig:setup}(b) for a schematic representation of this system.
The total force $F_x(y)$ on the medium particle for a given position $x$ of the probe is then
\bea
F_x(y) &=& -\ve A_a y  - \nabla_y U(x,y) \cr
&=&-\ve A_a y- A_s y + \la B(x - y) \cr
&=& -D\,(y - c_x)
\eea
with the notation $D = A +  \lambda B$, $A =  A_s + \ve A_a$, and
%$c(x) = c_{\lambda,\ve}(x) =\lambda\,D^{-1} B x$
$c_x = \lambda\,D^{-1} B x$ is the mechanical equilibrium position for the medium particle for a fixed probe position.

A more specific example in two dimensions takes
\[
A_s =\bbI\,,\quad
A_a =
\left( \begin{array}{cc}
0 & -1  \\
1 & 0
\end{array} \right),\quad
B = \left( \begin{array}{cc}
1 & b  \\
b & 1
\end{array} \right) \text{  for   } |b| < 1, \quad
K = \bbI\
\]
That describes a cloud of particles attached via a spring to the origin, subject to
a rotational force of strength $\ve$ and also harmonically coupled to the  probe.  In that case
\bea
D=
\begin{pmatrix}
 1+\la & \la b -\ve \\
\la b +\ve & 1 + \la
 \end{pmatrix},\quad
 c_x=\frac{\la}{ \zeta}
 \begin{pmatrix}
 (1 + \la (1-b^2) + \ve b) x_1 + (b + \ve) x_2 \\
 (1 + \la(1-b^2) - \ve b) x_2 + (b - \ve) x_1
 \end{pmatrix}
\eea
where $\zeta = (1 + \la)^2 - \la^2 b^2 + \ve^2.$  \\

The cloud dynamics is the overdamped diffusion $y_t$ at temperature $T$ with friction $\gamma$,
\bea
\ga\dot{y}_t=F_x(y_t) + \sqrt{2\ga T}\,\xi_t \label{langevin}
\eea
with standard $d-$dimensional white noise $\xi_t$.

For $x$ fixed the stationary density solving the Smoluchowski equation
\begin{equation}\label{eq:smol}
\nabla_y \cdot \Bigl[ \frac{F_x(y)}{\ga}\rho_x(y) -
\frac{T}{\ga}\,\nabla_y\, \rho_x(y) \Bigr] = 0, \qquad
\text{for all }~ y
\end{equation}
is the Gaussian density
\begin{equation}\label{stationary}
\rho_x(y) = {\cal N}\, e^{-\frac{1}{2T}\,(y - c_x)
\cdot \Ga \,(y - c_x)}
\end{equation}
where $\cal N = [\text{det}(\Ga) / (2\pi T)^d]^{1/2}$ is the normalization and
$\Ga$ is the (unique) positive symmetric matrix satisfying
\begin{equation}\label{Gamma}
D \Ga^{-1} + \Ga^{-1} D^\dagger = 2\bbI
\end{equation}
or, equivalently, the symmetric part of $\Ga D$ must equal $\Ga^2$; see Appendix \ref{clm}.
In particular, if $D$ is normal in the sense that it commutes with its transpose, $DD^\dagger = D^\dagger D$,
then the solution to~\eqref{Gamma} reads
$\Ga  = D_s = A_s + \lambda B$.   On the other hand, for non--normal $D$'s the matrix $\Ga$ does depend on $\ve$. Observe however also in \eqref{stationary} the temperature--dependence which is always of the Boltzmann-Gibbs form, so that the stationary density for the medium can  be seen as an equilibrium for the oscillator energy with ``spring constant'' $\Gamma$ and equilibrium position $c_x$.  The (nonequilibrium) $\ve-$dependence sits in $c_x$ and for non-normal $D$ also in $\Gamma$.
%(at least when $D$ is non-normal) and in $c_x$.
Note by comparing \eqref{hs}--\eqref{wowa} with \eqref{stationary},  that we already know that the nonequilibrium correction in the statistical force will also be temperature-independent.

Coming back to the above 2--dimensional example we find that $D$ is non-normal whenever $\ve b\neq 0$.  The stationary density is determined there by
\begin{equation}\label{2dim-stat}
\Gamma =
\frac{1}{\ka}
\left(
\begin{array}{cc}
1 + \la + \frac{\ve(\la b + \ve)}{1+\la} &
\la b
\\
\la b &
1 + \la - \frac{\ve(\la b - \ve)}{1+\la}
\end{array}
\right)
\end{equation}
with
$\ka = 1+ \bigl( \frac{\ve}{1+\la} \bigr)^2$.

The statistical force on the quasi-static probe follows from
$\nabla_x U(x,y)  = \lambda B(x - y) + K x$, and equals
\begin{equation}\label{linforce}
\begin{split}
f(x) &= -\int \nabla_x U(x,y)\,\rho_x(y)\,\id y
\\
&= \la B \,( c_x - x) -K x= -
\bigl( A + \frac 1 \la K B^{-1}D \bigr) \, c_x = - M x
\end{split}
\end{equation}
where $M = K + \la B -  \la^2 B D^{-1} B = K + \la (B^{-1} + \la A^{-1})^{-1}$. Of course the contribution of
$K$ is not statistical and was there as self-potential from the beginning. Note that the statistical force is temperature--independent. As for linear overdamped dynamics, rotational forces enter via asymmetric matrices and it is indeed useful to decompose $M= M_s + M_a$ with $M_s = (M+ M^\dagger)/2$, $M_a = (M-M^\dagger)/2$. The antisymmetric component $M_a$ quantifies the
induced rotational part and is of order $\ve \lambda^2.$

Continuing with the above 2-dimensional example, we have that the antisymmetric part equals
\begin{equation}
M_a = \frac{\ve {\la}^2 (1 - b^2)}{ (1 + \la)^2 - {\la}^2 b^2 + \ve^2}\, A_a
\end{equation}
and the symmetric part of $M$ obtains a second-order correction with respect to equilibrium,
\bea
M_s &=& K +\frac{\la}{\zeta}
\begin{pmatrix}
 1 + \la (1 - b^2) + \ve^2 & b(1+ \ve^2) \\
 b(1+ \ve^2) & 1 + \la (1 - b^2) + \ve^2
\end{pmatrix} \cr
&=&\frac 1{\zeta}
\begin{pmatrix}
 (1+\la)(1+2\la+\ve^2)- 2b^2 \la^2  & \la b(1+ \ve^2) \\
 \la b(1+ \ve^2) &  (1+\la)(1+2\la+\ve^2)- 2b^2 \la^2
\end{pmatrix}
\eea
Such a renormalization of the ``bare'' interaction constant due to coupling with environment is often called a ``Lamb shift''; here we see how it obtains a nonequilibrium contribution of order $O(\lambda \varepsilon^2)$ from the driving of the medium. In Fig.~\ref{fig:linear}(b) we plot the phase portrait of the statistical force under a specific choice of parameters.

The resulting motion of the probe depends of course on still other aspects of the medium and bath.  There will be friction and noise as further corrections to the statistical force, but for a quasi-static and macroscopic probe of mass $m$ we simply put
\[
m \ddot{x} = f(x) \;\,\;\;(\text{here } = -Mx)
\]
for its equation of motion.\\

To obtain yet another representation of the statistical force we observe (see Appendix \ref{clm}) that for \eqref{potential},
\begin{equation}\label{potent}
U(x,y) = \frac{1}{2}\,(y - c_x) \cdot D_s \, (y - c_x) +
c_x \cdot D_a (y - c_x) +
\frac{1}{2}\, x \cdot M_s \,x \n
\end{equation}
(with $D_a = \ve A_a$)
and the mean energy is
\begin{equation}
\langle U \rangle^x = \frac{d}{2}\,T + \frac{1}{2}\,x \cdot M_s \, x
\end{equation}
Combined with the medium's stationary entropy
\begin{equation}
\begin{split}
S(x) &= -\langle \log\rho_x \rangle^x
= \frac{d}{2} -
\frac{1}{2}\log \frac{\text{det}(D_s)}{(2\pi T)^d}
\end{split} \n
\end{equation}
we obtain the nonequilibrium free energy
\begin{equation}
\caF(x) = \langle U \rangle^x - T S(x)
= \frac{1}{2}\log \frac{\text{det}(D_s)}{(2\pi T)^d} +
\frac{1}{2}\, x \cdot M_s \,x
\end{equation}
The latter is to be compared with its equilibrium counterpart
\begin{equation}
\begin{split}
\caF_\text{eq}(x) &= -T\,\log \int e^{-\beta U(x,y)}\,\id y
\\
&= \frac{1}{2}\log \frac{\text{det}(D_s)}{(2\pi T)^d} +
\frac{1}{2}\,x \cdot M_s^{(0)} \, x
\end{split} \n
\end{equation}
with $M_s^{(0)} = K + \lambda[B - \la B D_s^{-1} B]$.
Note that from the Gibbs variational principle,
$\caF(x) \geq \caF_\text{eq}(x)$; their difference comes from
$M_s = M_s^{(0)} + O(\ve^2 \la)$.

As a consequence the statistical force \eqref{linforce} is manifestly a sum of two contributions,
\begin{equation}\label{stat-exact}
f(x) =-\nabla_x \caF(x) - M_a \, x =
-\nabla_x \langle U \rangle^x - M_a \, x
\end{equation}
The rotational force $-A_a y$ on the medium has been transformed into (i) a shift in the free energy which (still) determines the conservative component of the force, and (ii) an induced rotational force $-M_a x$. The total nonequilibrium correction to the statistical force on the probe is then given by
\bea
g(x) = -(M_s -M_s^{(0)})x - M_a x
\eea
According to the general theory the term $-M_a\,x$ corresponds to the
excess work of the rotational forces on the medium, at least close to equilibrium as we have argued in the previous Sections. We now check that within the present linear framework that is in fact \emph{exactly} (to all orders of $\ve$) verified (but the McLennan distribution \eqref{ml} is not exactly equal to \eqref{stationary}).

We need to calculate the excess work $V(x,y)$,
first when starting the medium from fixed position $y$ at fixed probe position $x$.  We follow the derivation in the first part of Appendix \ref{amc}.
The expected power of the total force $F_x(y)$ on the medium particle is equal to
\begin{equation} \label{eq:wz}
w_x(y) = \frac{1}{\ga}F_x(y)\cdot F_x(y) + \frac{T}{ \gamma} \nabla_y \cdot F_x(y)
\end{equation}
(see for example equation (III.5) in \cite{mcl} or Appendix \ref{amc}).  Then, following \eqref{defV},
\begin{equation}\label{qrt}
U(x,y) - \langle U\rangle^x + V(x,y) =
\int_0^\infty \bigl[ \langle w_x(y_t)|y_0=y\rangle^x - \langle w_x \rangle^x \bigr]\,\id t
\end{equation}
After some computation (see Appendix \ref{clm}) we find
the excess work due to the rotational force, starting from a fixed medium particle position $y$  given by
\begin{equation} \label{eq:Vz}
V(x,y) = \frac{1}{2}\,(y - c_x) \cdot
\Om \, (y - c_x) -\frac{T}{2}\,\text{Tr}\,(\Om \Ga^{-1})
- U(x,y) + \langle U \rangle^x
\end{equation}
where $\Om$ is a positive symmetric matrix such that
\begin{equation}\label{eq:Omega}
(D^{-1})^\dagger \Om + \Om D^{-1} = 2\bbI
\end{equation}
(When $D$ is a normal matrix, then
$\Om^{-1} = (D^{-1})_s$, and $\Om = \Ga - \Ga^{-1} D_a^2$.)

The averaged excess work \eqref{wex}, when the probe is shifted from $x \to x+\id x$ is
\[
\dbar W^\text{ex}(x) = \int\id y\,\rho_x(y)\,[V(x+\id x, y)- V(x,y)]
\]
We have, from \eqref{eq:Vz}, to linear order in $\id x,$
\bea
\dbar W^\text{ex}(x) &=& \nabla_x \langle U \rangle^x \cdot \id x - \langle \nabla_x  U \rangle^x \cdot \id x \cr
&=& M_s x \cdot \id x - M x \cdot \id x \cr
&=& -M_a x\cdot \id x
\eea
Thus the excess work dissipated by the medium due to the rotational force is equal to the work done on the probe by the rotational component of the statistical force when the probe position is shifted by an amount $\id x.$\\

For the two-dimensional example we find
\[
\Om =
\left(
\begin{array}{cc}
1 + \la + \frac{\ve(\la b + \ve)}{1+\la} &
\la b
\\
\la b &
1 + \la - \frac{\ve(\la b - \ve)}{1+\la}
\end{array}
\right)
\]
and by comparing with~\eqref{2dim-stat} we check the relation
$\Om = \ka\, \Ga$ (moreover, $\Ga \Om = D^\dagger D$) which means that the stationary distribution of the cloud is given by an exact variant of the McLennan ensemble
$\rho_x(y) \propto \exp\,[-(U+V)(x,y) / T_\text{eff}]$, with the modified potential
$U + V$ and the ``renormalized'' temperature
$T_\text{eff} = \ka T  = T + O(\ve^2)$. As a consequence the two-dimensional model satisfies the exact generalized Clausius relation
$\dbar Q^\text{ex} = T_\text{eff}\, \id S$ (with respect to all possible thermodynamic transformations) where
$S(x) = -\langle \log\rho_x \rangle^x$ is the stationary (Shannon) entropy.

\section{Realizing minimum entropy production}\label{min}

In that same linear regime, statistical forces should reflect the tendency of the compound system (probe plus nonequilibrium medium) to reach the condition of minimum entropy production rate (MINEP), \cite{scholar}, valid for close-to-equilibrium media with degrees of freedom that are even under kinematic time-reversal. We show now that the opposite also holds giving a third proof of \eqref{ge}: requiring MINEP implies that the work needed to move the probe over $\id x$ equals the change in equilibrium free energy plus excess work done by the nonconservative forces on the medium to relax from the old stationary condition $\rho_x$ to the new one described by $\rho_{x+\id x}$.

\subsection{Minimal nonequilibrium free energy}
Before we go to the actual application it is useful to derive an alternative (but equivalent) formulation of the minimum entropy production principle in terms of a nonequilibrium free energy functional.

Suppose we have states $\sigma$ (they will be the states $\sigma = (x,\eta)$ of our compound system) and probability distributions $\mu$ on them.
There is a driven process $\sigma_t$ that satisfies the condition of local detailed balance.  There is a unique stationary distribution $\rho$ and obviously there is a trivial variational principle $s(\mu \rel \rho) \geq 0$ with equality only if $\mu=\rho$ in terms of the relative entropy
$s(\mu \rel \rho) = \sum_\sigma \mu(\sigma)\,\log\,[\mu(\sigma)/\rho(\sigma)]$
(for simplicity we take here finite irreducible Markov processes).  That variational formula starts being useful if $\log \rho(\sigma)$ has a (thermo)dynamical meaning. That is certainly the case at equilibrium but also near equilibrium where $\rho=\rho^{\text{ML}} + O(\varepsilon^2)$ in \eqref{ml} is expressed in terms of energy and excess work.  If we find $\mu$ that minimizes
\begin{equation}\label{rel}
s(\mu \rel \rho^{\text{ML}}) =\sum_{\sigma} \mu(\sigma) \log \mu(\sigma) + \beta\sum_{\sigma} \mu(\sigma)\,(U+V)(\sigma) +
\log{\cal Z} \geq 0
\end{equation}
then (obviously) $\mu = \rho^{\text{ML}}$ and $\mu =\rho + O(\varepsilon^2)$ is a perfect linear order approximation to the true stationary distribution.  In the expression \eqref{rel} we recognize the time-integrated entropy production for the process relaxing from $\mu$ {\it versus} from the McLennan distribution.  That is because there $\log {\cal Z} = -\beta\sum_\sigma(U+V)(\sigma) \,\rho^{\text{ML}}(\sigma) + S(\rho^{\text{ML}})$.
In that sense we do exactly what the minimum entropy production is doing, and in fact by requiring
$\id/\id t\,s(\mu_t \rel \rho^{\text{ML}})_{t=0} \leq 0$ we would even recover it in its usual
instanteneous version, \cite{scholar}.

Let us still rewrite~\eqref{rel} using the  variational nonequilibrium free energy functional~\cite{Naoko}
\begin{equation}\label{dff}
{\cal F}_\text{neq}(\mu) = \sum_\sigma(U+V)(\sigma) \,\mu(\sigma) - T\,S(\mu)
\end{equation}
At stationarity $\mu=\rho$ it coincides with  the
usual free energy functional
${\cal F}_\text{neq}(\rho) = {\cal F}(\rho) =   \langle U \rangle - T\,S(\rho)$, since $\langle V \rangle = 0$, and
${\cal F}_\text{neq}(\rho^{\text{ML}}) = -T\log {\cal Z}$.  Furthermore, the positivity in \eqref{rel} gives the variational principle
\begin{equation}
\label{av}
{\cal F}_\text{neq}(\mu) \geq {\cal F}_\text{neq}(\rho^{\text{ML}})
\end{equation}
with equality for $\mu = \rho^\text{ML} = \rho + O(\varepsilon^2)$. That is the free energy version of MINEP:  correct to first order $\varepsilon$ the stationary distribution is the one that has lowest (nonequilibrium) free energy.
Note also that
${\cal F}(\rho) = -T\,\log Z + O(\ve^2)$, and
$\sum_\sigma \mu(\sigma) V(\sigma) = O(\ve^2)$ whenever $\mu / \rho = 1 + O(\ve)$.
To the best of our knowledge that formulation is new and especially useful for work-considerations as arise in the context of the present paper.

\subsection{Application to the probe--medium system}
The previous principle will be applied to the compound system of probe plus medium.  We make however an additional simplifying assumption,
that we can characterize the statistical force at probe position $x^*$ by finding the constant force $B$ so that when applying $B$ to the probe it actually relaxes to position $x^*$ as unique attractor and fixed point.  Then,  $f(x^*) = -B$; in other words, $B$ exactly cancels the statistical force at steady position $x^*$.  We now consider the modified dynamics with that additional constant force $B$ on the probe and we require that
$\rho(x,\eta) = \delta(x-x^*)\,\rho_{x^*}(\eta)$ is the stationary distribution (always taken to first order in $\epsilon$).  That requirement will be implemented by the free energy principle (or MINEP)  \eqref{av}.\\

Let us take as test-distribution $\mu(x,\eta) = \delta(x-z)\rho_z(\eta)$ that would put the probe at $z$ and take the McLennan distribution $\rho_z$ for the medium.  We now write for that choice
$\caF_\text{neq}(\mu)  = \caF_\text{neq}(z)$ which from \eqref{dff} becomes
\[
\caF_\text{neq}(z) = -B\cdot z + \langle U(z,\eta) \rangle^z  - TS(\rho_z) + \int_{\gamma: z\rightarrow x^*}\,\dbar W^\text{ex}
\]
where the last line-integral gives the excess work when moving from $z$ to $x^*.$ The principle \eqref{av} tells us that
\begin{multline}
-B\cdot z + \langle U(z,\eta) \rangle^z  - TS(\rho_z) + \int_{\gamma: z\rightarrow x^*}\,\dbar W^\text{ex}
\\
\geq -B\cdot x^* + \langle U(x^*,\eta) \rangle^{x^*}  - TS(\rho_{x^*}) = -B\cdot x^* + {\cal F}_\text{eq}(x^*)
\end{multline}
where we inserted (correct to first order) the equilibrium free energy.  We insert $z= x^* + \id x^*$ for small deviations around the attractor and find at the minimum
\[
\id{\cal F}_\text{eq}(x^*)
 - \dbar W^\text{ex}(x^*) = B\cdot \id x^*
\]
which is again the sought result as $B=-f(x^*)$.
Supposing there is a unique $x^*$ at which $f(x^*)=0$, that point is characterized by minimizing the nonequilibrium free energy  $\caF_\text{neq}.$

For example, when two reservoirs are in mechanical contact, separated by a piston, the piston will move to equalize the two pressures but the pressure is not just the derivative of the equilibrium free energy; one will need to estimate the change in excess work under variations of the piston position.  More specifically, consider a gas in a vessel divided into two compartments with volumes $\Lambda_1 + \Lambda_2 = \Lambda$ via a movable piston, under isothermal conditions. If we start ``stirring'' the gas in compartment $1$, the piston gets moving to continuously decrease the nonequilibrium free energy
\begin{equation}
{\cal F}_\text{neq}(\Lambda_1,\Lambda_2) = {\cal F}_1(\Lambda_1) + {\cal F}_2(\Lambda_2) -
\int^{\Lambda_1} \dbar W_1^\text{ex}
\end{equation}
until it attains minimum. The latter of course corresponds to equalizing pressures
$P_1 = P_2$ with $P_1$ obtaining a nonequilibrium correction,
$P_1 = -\id {\cal F}_1 / \id \Lambda_1 + \dbar W^\text{ex} / \id \Lambda_1$.

Note that here again we have considered the physical context of even degrees of freedom for the medium.  As is known, the minimum entropy production principle does not apply with velocity degrees of freedom; see e.g. \cite{scholar}.  Yet, the mathematics and the formal arguments as presented above remain of course valid as such, although in that case of non-even degrees of freedom without a direct physical interpretation.

\section{Measuring dynamical activity}\label{2ch}
Dynamical activity measures the time-symmetric current or the number of transitions in a given space-time window.  It is the change in that activity when perturbing the system, or the relative activity when comparing different transition paths, that matters in response theory \cite{frenesy}.  In fact, also for detailed balance dynamics, the dynamical activity appears important for understanding aspects of jamming and glass transitions \cite{vW,chan}.  However,  dynamical activity is difficult to access directly. Here we look into a toy example demonstrating how nonequilibrium statistical forces could be used to (indirectly) measure the relative activity, at least in the case of a simple state-space geometry. \\

\begin{figure}[ht]
 \centering
 \includegraphics[width=7 cm]{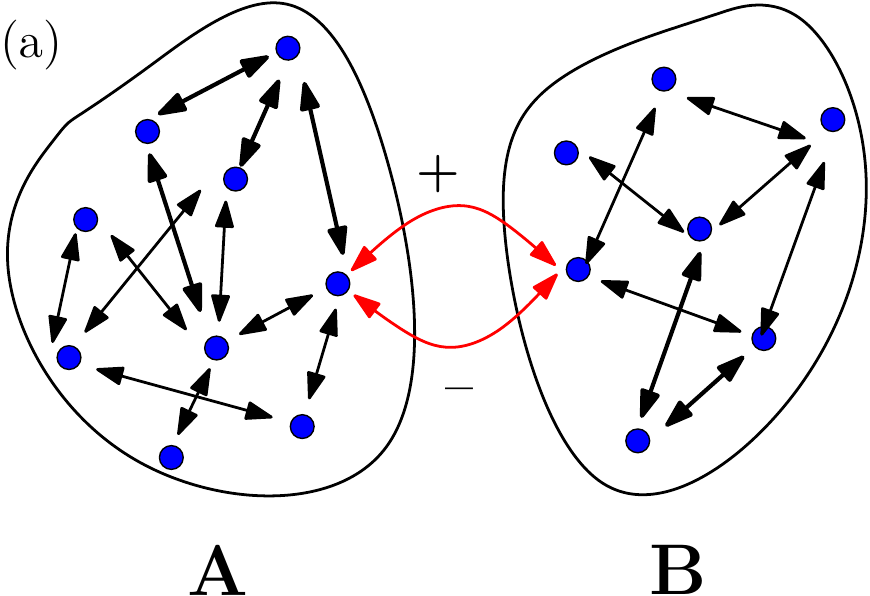} \hspace*{0.5 cm}  \includegraphics[width=7 cm]{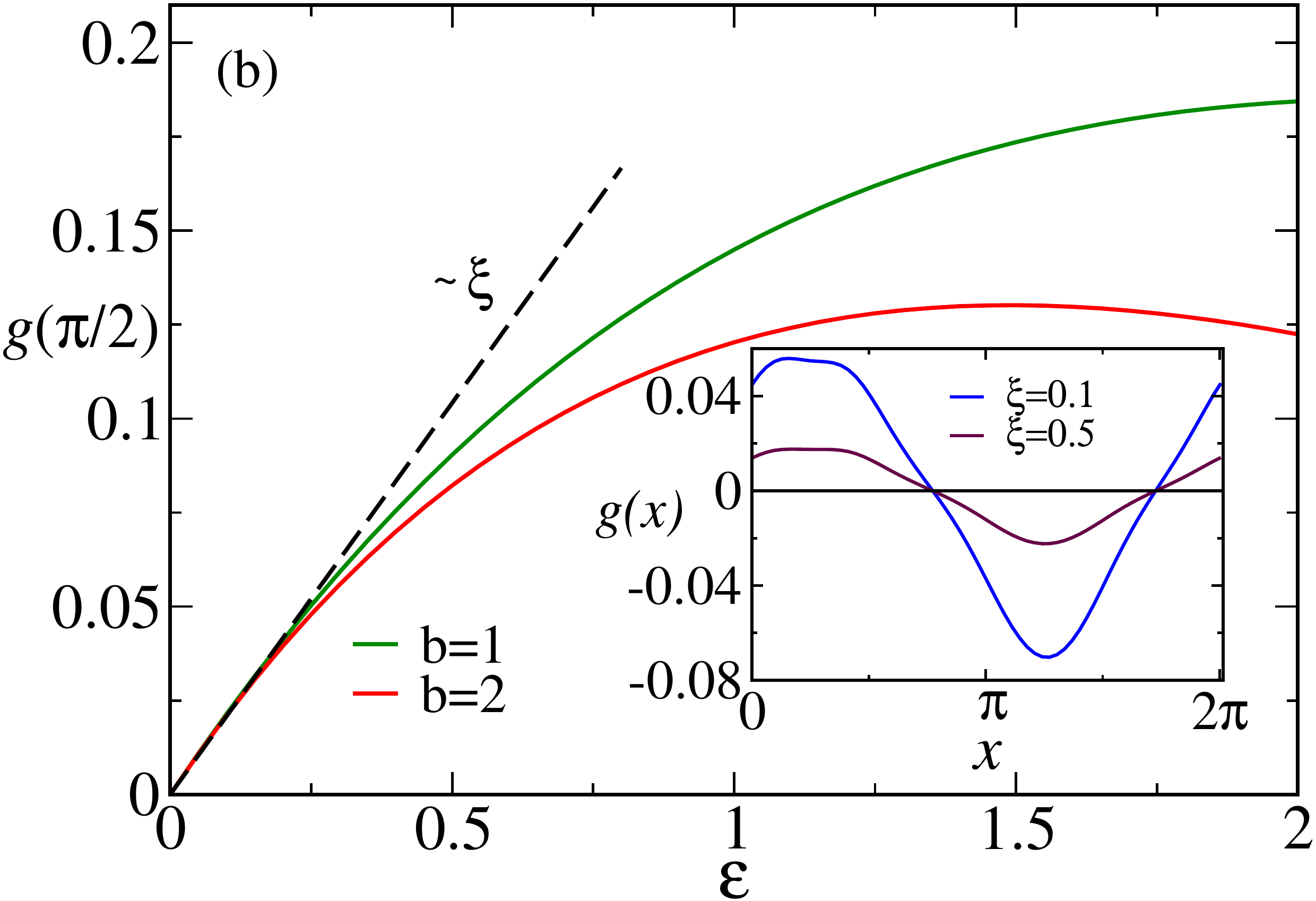}
 % frenesy_met.pdf: 595x842 pixel, 72dpi, 20.99x29.70 cm, bb=0 0 595 842
 \caption{Frenometer: (a) Schematic representation of the two types of configurations connected through a two-channel transition path.  (b) The nonequilibrium correction to the statistical force for the toy model (for a fixed $x=\pi/2.$) as a function of the driving $\ve$ through the bottleneck for two different values of $b=1,2.$ and $\xi=0.1.$  We can read the relative dynamical activity in equilibrium from the slope. The inset shows the same correction $g(x)$ as a function of $x$ for a fixed $\ve=0.2$ and different $\xi=0.1,0.5.$ Here $b=1.$}
 \label{fig:fren_met}
\end{figure}

Assume we have an \emph{equilibrium} system of noninteracting particles the configuration space of which splits into two parts, $A$ and $B$, connected through a two-channel bottleneck only; see Fig.~\ref{fig:fren_met} ---  we call them the $+$ and $-$ channel.  We want to find out which of the two channels is more ``open'' in terms of their relative dynamical activities. The idea is to connect this question to the problem of how statistical forces respond to switching on a weak nonequilibrium force in the bottleneck.\\

The bottleneck consists of a pair of (single-particle) transitions
$A\ni \si_A \stackrel{+,\,-}{\leftrightsquigarrow} \si_B \in B$ with rates
$k^o_\pm(\si_A,\si_B) = \ga^o_\pm\,\exp\,(\frac{\be}{2}[U(\si_A) - U(\si_B)])$ respectively
$k^o_\pm(\si_B,\si_A) = \ga^o_\pm\,\exp\,(\frac{\be}{2}[U(\si_B) - U(\si_A)])$. The rest of the system is arbitrary up to that the transitions satisfy detailed balance with potential $U$ and $k^o(\eta,\eta') = 0$ whenever $\eta$, $\eta'$ do not belong either both to $A$ or both to $B$.   We are to determine the dynamical activities
$D_\pm$ defined as the mean equilibrium frequency of transitions along the channels $\pm.$
From detailed balance,
\bea
\frac{D_+}{D_-} = \frac{\rho^{\text{eq}}(\sigma_A) k_+^o(\si_A,\si_B)+\rho^{\text{eq}}(\sigma_B) k_+^o(\si_B,\si_A)  }{\rho^{\text{eq}}(\sigma_A) k_-^o(\si_A,\si_B)+\rho^{\text{eq}}(\sigma_B) k_-^o(\si_B,\si_A)}
= \frac{\gamma_+^o}{\gamma_-^o}
\eea
Such a relative dynamical activity and its further dependence on nonequilibrium parameters is an example of what we call more generally frenetic aspects, to contrast it with entropic features.  In that way, the set-up in Fig.~\ref{fig:fren_met} represents a frenometer as we now show.

Now enters the interaction with a probe. Therefore, we let the energy $U$ also depend on the position $x$ of the probe.
The equilibrium statistical force on the probe is derived from the free energy.  The partition function is the sum over all states,
$Z_{A,B} = \sum_{\eta \in A,B} e^{-\be U(x,\eta)}$ and  that force equals
\begin{equation}
\begin{split}
f_\text{eq} &= T\,\nabla_x\log(Z_A + Z_B)
\\
&= \rho_A f_A + \rho_B f_B
\end{split}
\end{equation}
where $f_{A,B} = T\,\nabla_x \log Z_{A,B}$ is the mean force from $A$ and $B$, respectively, and
$\rho_{A,B} =  Z_{A,B} / (Z_A + Z_B)$ is the proportion of time the equilibrium system spends in each compartment.\\
In order to measure the relative activity $D_+ / D_-$, we drive the system out of equilibrium by applying a local nonpotential force which modifies the transition rates in the bottleneck to
\begin{align}
k_\pm(\si_A,\si_B) &= \ga_\pm e^{\frac{\be}{2}[U(x,\si_A) - U(x,\si_B) \pm \varepsilon]} \n
\\
k_\pm(\si_B,\si_A) &= \ga_\pm e^{\frac{\be}{2}[U(x,\si_B) - U(x,\si_A) \mp \varepsilon]}
\end{align}
A possible  $\varepsilon-$dependence of the kinetic factors $\ga_\pm$ is allowed but irrelevant for linear order calculations;
see below in \eqref{bb} and the dependence on the parameter $b$ in Fig.~\ref{fig:fren_met}. %and also \cite{PRL-attempt}
We show next that the nonequilibrium correction to the statistical force gives information about $D_+ / D_-$.

An immediate effect of turning on the drive is a redistribution of the particles, described to linear order in $\varepsilon$ by the McLennan ensemble \eqref{ml}. For the $V(x,\eta)$ there we need the work performed by the applied force along those parts of the relaxation trajectories that pass the bottleneck.  All trajectories, say originating from part $A,$ have to pass through the ``port'' $\si_A$ in order to access the bottleneck, and since no other transitions contribute to dissipated work but the two special channels $\pm$, $V(x,\eta)= V_{A,B}$ is constant inside both $A$ and $B$.  A calculation to linear order (Appendix \ref{amc}) yields
\begin{equation}\label{cm}
V_A - V_B = \varepsilon \xi \,,\quad
\xi = \frac{D_+ - D_-}{D_+ + D_-}
\end{equation}
supplied with the normalization condition
$Z_A V_A + Z_B V_B = 0$. The difference $V_A - V_B$ can be detected as a nonequilibrium correction to the statistical force acting on the external slow particle. By formula~\eqref{11} and since $h_x(\eta) = -\beta V(x,\eta) + O(\varepsilon^2)$, that correction equals 
\begin{equation}
\begin{split}
g(x) &= \langle \nabla_x V \rangle^{x,\text{eq}} = -\sum_{\eta\in A} V(x,\eta) \nabla_x \rho_x^{\text{eq}}(\eta) -\sum_{\eta\in B} V(x,\eta) \nabla_x \rho_x^{\text{eq}}(\eta)
%= -\sum_\eta \nabla_x \rho_x^o(\eta)\,V_x(\eta)
\\
&= -V_A \nabla_x \left( \frac{Z_A}{Z_A + Z_B} \right) -
V_B \nabla_x \left( \frac{Z_B}{Z_A + Z_B} \right)
\\
&= -\frac{\varepsilon \,\xi}{2} \nabla_x \left( \frac{Z_A - Z_B}{Z_A + Z_B} \right) + O(\varepsilon^2)
\end{split}
\end{equation}
In terms of the (equilibrium) occupations and statistical forces associated with $A$ respectively $B$, the nonequilibrium correction to the statistical force takes the form
\begin{equation}
g = \varepsilon \xi \be  (f_B - f_A)\, \rho_A \rho_B \label{conf} + O(\varepsilon^2)
\end{equation}
Hence, the channel-asymmetry factor $\xi$, characterizing the relative importance (in terms of dynamical activity) of the two channels, can be evaluated from the first order correction of the statistical force.  It determines the slope in Fig.~\ref{fig:fren_met}(b) in the close-to-equilibrium dependence of the statistical force on the nonequilibrium amplitude $\varepsilon$ given that we know the equilibrium values $f_{A,B}$ and $\rho_{A,B}.$\\

For illustration we take a simple toy system where both $A$ and $B$ are two three-state rotators with states $\eta_{A,B}=-1,0,1$ with bottleneck states $\sigma_A=1_A$ and $\sigma_B=1_B$ connected by two channels.  The probe is connected to the rotators via interaction energy
\bea
U(x,\eta_\alpha) = \delta_{\alpha,A}[\eta \sin x+ 2\eta^2\cos x] + \delta_{\alpha,B}[\eta \cos x+ 2\eta^2\sin x]  \n
\eea
Note that the specific form of the interaction potential is not of any particular relevance; the above choice just avoids special symmetries. We assume that the drive affects the reactivity of the $+$ channel
\begin{equation}\label{bb}
\gamma^\ve_+ = \gamma^o_+(1+b |\ve|)
\end{equation}
for some constant $b.$
The nonequilibrium correction to the force $g$ as a function of the drive $\ve$ for a fixed probe position $x$
is shown in Fig. \ref{fig:fren_met}(b); the slope of the curves close to $\ve=0$ is determined by the channel asymmetry factor $\xi$ confirming \eqref{conf}. But there is more:  the second order is able to pick up the $\ve-$dependence (parameter $b$) in the channel reactivities, invisible to linear order.
That is in line with the analysis of higher order effects in the response formalism in \cite{pccp}.
We  can thus measure the changes in time-symmetric aspects of the medium due to its nonequilibrium condition, from observing the probe's motion.

\section{Conclusion}

We have discussed in detail how the statistical force of a medium becomes modified when the medium is weakly driven out of equilibrium. Independent of the nature of the driving, the systematic nonequilibrium force is intimately related to the steady state thermodynamics of the medium as governed by the (slow) motion of an attached probe. In this way, a simple measurement on the probe can reveal the excess work of driving forces in the medium, which is hard to be measured directly as it requires to distinguish a rather tiny effect against an omnipresent dissipative background. It was demonstrated how this result emerges both thermodynamically (via a generalized Clausius relation) and statistical-mechanically (via the McLennan nonequilibrium ensemble). We have also formulated a variational principle for the
point attractors of the macroscopic probe in terms of a nonequilibrium generalization of the free energy which realizes the minimum entropy production principle.
Finally, we have shown how to set up a ``frenometer,'' using the statistical force to measure relative and excess dynamical activities.  That can be important as it adds operational meaning to that time-symmetric
variant of current which is known to be important for nonequilibrium response theory.

From a more general perspective, the analysis of statistical forces poses a complementary (mechanical) problem to the (calorimetric) problem of heat exchange between the medium and its thermal environment, which can be quantified via nonequilibrium heat capacities. Establishing quantitative relations between both sectors remains a relevant and nontrivial problem of steady state thermodynamics.

%------------------------------------------
\appendix

\section{Excess work}\label{amc}

We start with a brief review of the McLennan ensemble for the purpose of this paper; see~\cite{mac,mcl} for more details.   That ensemble summarizes the static fluctuations in the linear regime around a detailed balance dynamics.  One can get the linear response relations, including Kubo and Green-Kubo formul{\ae} directly from it. Interestingly however, the McLennan ensemble can be obtained from physically specified quantities, and therefore can be formulated even without detailing the dynamics.  The most elegant and physically direct way to obtain that ensemble is in \cite{naokom} and starts from a perturbation expansion of an exact fluctuation symmetry for the irreversible entropy fluxes.
The main player there is the excess work whose meaning is already visible from Fig.~\ref{fig:setup}(b).
The excess work is associated to a force $G$ which is doing work on the medium and that is dissipated in the heat bath. That force $G$ can be the total force or only its non-conservative part or even something else.  

To be specific we imagine an overdamped diffusion process $y_t$ much as in \eqref{langevin},
\[
\gamma \,\dot y_t =  -\nabla U(y_t) + F^a + \sqrt{2\gamma T}\,\xi_t
\]
where we split up the total force into a conservative part with potential $U$ and $F^a$ stands for the driving force.  (There is no need to be precise about this splitting of the total force for defining the McLennan ensemble.)
The expected current is $j_\mu = \frac 1{\gamma} F \mu - \frac T{\gamma}\nabla\mu$ when the distribution over $y$ is $\mu$ where the total force is $F =  -\nabla U(y_t) + F^a$.

The instantaneous mean power associated to $G$ is%work
\[
W^G(\mu) = \int G(y) \cdot j_\mu(y) \, \id y
\]
We thus have $W^G(\mu) = \int \id y \, w^G(y) \,\mu(y)$ and
\[
w^G(y) = \frac 1{\gamma} G(y)\cdot F(y) +\frac T{\gamma} \nabla\cdot G(y)
\]
is the dissipated power when in state $y$.    If $G=F$ the total force, the last identity is recognized in \eqref{eq:wz} (with total force also still depending on the probe position $x$).
Note that $w^G$ is linear in $G$ so that the power (and excess) is additive in the force $G$.  To go to the excess we need to subtract the stationary dissipative power and integrate over time to get the excess work $V^G$:
\begin{equation}\label{defV}
V^G(y) = \int_0^{\infty}\,\id t\,\left[\langle w^G(y_t)|y_0=y\rangle - \langle w^G\rangle\right]
\end{equation}
For example, by taking $G=-\nabla U$, we get
 \[
 w^{G}(y) = -\frac 1{\gamma} \nabla U(y)\cdot F(y) - \frac T{\gamma} \Delta_y U(y) = -L \, U
 \]
for backward generator $L$,
\bea
L = \frac 1 \gamma F\cdot \nabla  + \frac {T}{\gamma} \Delta \label{ell}
\eea

Formally in \eqref{defV}, $V^G= -L^{-1} w^G$ (with $\langle V^G \rangle=0$)  is the result of acting with the pseudo-inverse $L^{-1}$, and therefore the excess work by the conservative force  equals $U(y) - \langle U\rangle$ when relaxing from $y_0=y$.

The $V$ in the McLennan ensemble starting in \eqref{ml} and throughout the paper is the excess work associated to the driving force $G = F^a,$ or  $V=  V^{F^a}$.  Formula \eqref{qrt} gives the excess work as defined in the McLennan-ensemble for the total force $F_x(y) = \varepsilon A_a y - \nabla_y U(x,y)$, including the conservative part.  The reason to include there that conservative part is the simplicity of the Ansatz \eqref{eq:Vz} which is further discussed in the next Appendix.\\

A second computation of excess work (for jump processes) leads to the result in formula~\eqref{cm}.  We already mentioned there that $V(\eta) = V(\sigma_A)$ when $\eta\in A$, and similarly $V(\eta) = V(\sigma_B)$ when $\eta\in B$. That is because the only transitions with irreversible dissipation are those through the two channels at the bottleneck. When computing the excess work $V(\eta)$ (now only due to the nonconservative forces) in general we must look at the expected excess dissipation, and thus here
\[
V(\eta) = \varepsilon N_\eta(A\stackrel{+}{\rightarrow} B) - \varepsilon N_\eta(A\stackrel{-}{\rightarrow} B) + \varepsilon N_\eta(B\stackrel{-}{\rightarrow} A) - \varepsilon N_\eta(B\stackrel{+}{\rightarrow} A)
\]
where the $N_\eta$ are the expected total number of transitions when starting the equilibrium process in $\eta$.  Those expected number of transitions are determined by the transition rates and the expected number of visits:
\[
V_A-V_B = \varepsilon(\gamma_+ - \gamma_-)\int_0^{+\infty} \{k_{AB}\,[p_t(A,A)-p_t(B,A)] + k_{BA}\,[p_t(B,B)-p_t(A,B)]\}\,\id t
\]
where the $p_t$ are transition probabilities and the
$k_{AB} = k_{BA}^{-1} = \exp\, [U(\sigma_A) - U(\sigma_B)]/2$.  The rest of the computation uses detailed balance to reduce the case to that of a two state model with two channels. The approach to equilibrium is exponentially fast with rate $ r = (\gamma_+ + \gamma_-)\,[k_{AB} + k_{BA}]$.  Integrating over time
$\exp\,[-rt]$ gives the required formula \eqref{cm}.

% ----------------------------------------------------
\section{Computations for the linear model}\label{clm}

In this Section we give the explicit computations leading to the statistical force and excess work for the linear model studied in Section \ref{line}.

Under stationarity the position of the cloud particle fluctuates around the average $c_x$ for a fixed probe postition $x.$  For this linear system, the stationary density  $\rho_x(y)$ then must be a Gaussian of the form Eq.~\eqref{stationary},
\bea
\rho_x(y) = {\cal N} \exp \Bigl[-\frac 1{2T} (y-c_x) \cdot \Gamma (y-c_x) \Bigr] \n
\eea
where $\Gamma$ is a positive symmetric matrix which is to be determined from the Smoluchowski equation~\eqref{eq:smol} with
$F_x(y)=-D(y-c_x).$ Then,
\bea
\nabla_y ~\rho_x(y) &=& - \frac 1{T}\, \Gamma (y-c_x)~ \rho_x(y) \cr
\Delta_y ~\rho_x(y) &=& \frac 1{T^2}\, \Gamma (y-c_x) \cdot \Gamma (y-c_x)~\rho_x(y) - \frac 1{T}\, \text{Tr}[\Gamma]~ \rho_x(y) \n
\eea
and $\nabla_y \cdot F_x(y) = - \text{Tr}[D].$ Substituting the above in Eq. \eqref{eq:smol} we get that the symmetric matrix $\Gamma$ must satisfy the following relations,
\bea
(y-c_x)\cdot D^\dagger \Gamma (y-c_x) &=& (y-c_x) \cdot \Gamma^2 D(y-c_x) \cr
\text{and}~~~~ \text{Tr}[\Gamma] &=& \text{Tr}[D] \label {eq:gam_D}
\eea
The first equality demands that the symmetric part of  $D^\dagger \Gamma$ is equal to $\Gamma^2$ which is expressed by Eq.~\eqref{Gamma}. Multiplying Eq.~\eqref{Gamma} from the right with $\Gamma$, one gets
$D + \Gamma^{-1} D^\dagger \Gamma = 2 \Gamma.$ Taking the trace on both sides leads to the second equation above.

Next we detail  the computational steps leading to the alternative form of the energy $U(x,y)$  \eqref{potent}. From \eqref{potential} we have
\bea
U(x,y) = \frac 12 y \cdot D_s y - \la y \cdot B x + \frac 12 x \cdot (\la B+ K ) x \n
\eea
Replacing $\la  B x $ by $D c_x$ (from the definition of $c_x$) in the second term and performing a few steps of algebra we have,
\bea
U(x,y) = \frac 12  (y-c_x) \cdot D_s (y-c_x) + c_x \cdot D_a (y-c_x) - \frac 12 c_x \cdot D_s c_x + \frac 12 x \cdot (\la B+ K ) x \label{eq:U_int}
\eea
where we have used $c_x \cdot D_a c_x =0$ for the antisymmetric matrix $D_a.$ Now, again using the definition of $c_x$
\bea
c_x \cdot D_s c_x = \frac 12 \la^2 x \cdot B ((D^{-1})^\dagger + D^{-1}) B x \n
\eea
Substituting this in \eqref{eq:U_int} leads to \eqref{potent} where
\bea
M_s = K+ \la B - \frac 12 \la^2 B [(D^{-1})^\dagger + D^{-1}] B \n
\eea

%We give the explicit computations for the excess work in context of the linear model studied in Sec. \ref{line}.

Finally, the excess work when the medium particle starts from a fixed position $y$ for a given fixed probe position $x$ is related to the power of the driving force through
\bea
L (U(x,\cdot) + V(x,\cdot))(y) = -w_x(y) + \langle w_x \rangle
\label{eq:gen_w}
\eea
where $L$ is the backward generator for the cloud particle as in \eqref{ell}.
From Eq. \eqref{eq:wz} we have
\bea
w_x(y) &=& \frac{1}{\ga}F_x(y)\cdot F_x(y) + \frac{T}{ \gamma} \nabla_y \cdot F_x(y)  \cr
&=& \frac 1 \gamma D(y-c_x)  \cdot D(y-c_x) - \frac{T}{ \ga} \text{Tr}[D] \cr
\text{and}~~~\langle w_z(y) \rangle &=& \frac {T} {\gamma} \text{Tr}[ D^\dagger D \Gamma^{-1}] - \frac{T}{ \ga} \text{Tr}[D] \label{eq:wzav}
\eea

The excess work  performed by the total force on the medium particle must be of the form \eqref{eq:Vz} for some symmetric matrix $\Om$  because there is no force, $F_x(y) = 0$ when $y=c_x$ and the work must be symmetric in $y-c_x$.  We will find $\Omega$ from requiring \eqref{eq:gen_w}. The left-hand side of \eqref{eq:gen_w} can be calculated using~\eqref{eq:Vz},
\bea
\nabla V(x,y) = \Om (y-c_x), \;\; \text{and} \;\; \Delta_y V(x,y) = \text{Tr}[\Om]
\eea
Hence,
\bea
L V(x,y) = -\frac 1 \gamma D(y-c_x) \cdot \Om (y-c_x)  + \frac {T} {\gamma} \text{Tr}[\Om ] \n
\eea
Demanding \eqref{eq:gen_w}, we must have
\bea
(y-c_x)\cdot D^\dagger \Om (y-c_x) &=& (y-c_x) \cdot D^\dagger D(y-c_x) \cr
\text{and}~~~~~~ \text{Tr}[\Om] &=& \text{Tr}[ D^\dagger D \Gamma^{-1}] \n
\eea
Similar to \eqref{eq:gam_D} above, the first equation states that the symmetric part of $D^\dagger \Om$ is equal to $D^\dagger D$ which results in Eq. \eqref{eq:Omega}. The second equality above follows from there using Eq. \eqref{Gamma}
because $D^\dagger \Omega \Gamma^{-1} + \Omega D\Gamma^{-1} = 2 D^\dagger D \Gamma^{-1}$ of which we can take the trace with left-hand side giving 2Tr$[\Omega]$.

% ------------------------------------------------------------

\end{document}